# Quantum Entanglement and Special Relativity: A Possible Solution to the Paradox?

Yoram Kirsh[1]

**The Open University of Israel**[2]

**ABSTRACT**

Demonstrations of quantum entanglement which confirm the violation of Bell's inequality indicate that under certain conditions action at a distance is possible. This consequence seems to contradict the relativistic principle of causality, which asserts that an effect never precedes its cause, in any reference frame. By analyzing a numerical example of Bell's experiment with entangled pairs of photons, we show how observers in two inertial reference frames can disagree about the causality relation between two events. One observer claims that event 1 is the cause of event 2, while the other claims that event 1 is the result of event 2. The solution we suggest to the paradox is that in entangled systems, one can find pairs of "entangled events" which have symmetrical causality relations. Each of the events can serve as a cause or as an effect, depending on the frame of reference in which they are observed.

## 1. The EPR Paradox

Recently, several independent "loophole-free Bell violation" experiments were reported (e.g. Hensen et al., 2015, Giustina et al., 2015, Shalm et al., 2015). These works, as well as previous demonstrations of quantum entanglement since the pioneering work of Freedman and Clauser (1972), have shown that under certain conditions, action at a distance between entangled systems is possible. A straightforward interpretation of the results is that faster-than-light (FTL) or *superluminal* communication is feasible. This consequence may seem to contradict an established tenet of Special Relativity (SR) that claims that FTL transfer of information violates causality and is therefore impossible. The purpose of this article is to examine in detail where the conflict arises and to propose a possible solution to the paradox.

In fact, the paradox was already hidden in the EPR gedanken-experiment (Einstein, Podolsky, Rosen, 1935). If two identical particles, *A* and *B*, move in opposite directions after a brief interaction which ensured that $\mathbf{v}_B = -\mathbf{v}_A$, and if at a specific moment, $t_0$, when *A* and *B* are far apart, the position of *A* and the momentum of *B* are measured, one can know both the position and the momentum of *A* at $t_0$, since $\mathbf{p}_A = -\mathbf{p}_B$. The authors claimed that this thought experiment contradicted Heisenberg's uncertainty principle, and consequently that quantum mechanics is incomplete.

---

[1] yoramk@openu.ac.il
[2] 1 University Road, P.O.B 808, Raanana 43107, Israel

In addition to this explicit criticism against QM, another claim can be based on the EPR thought experiment, concerning an inconsistency between QM and SR. Suppose we perform a very accurate measurement of the magnitude of the momentum of *A* and find that it is $p_A$. According to the uncertainty principle, the measurement changes the state of *A* so that the uncertainty in its location becomes infinite, or very large (since $\Delta p \times \Delta x \approx \hbar$). But since *A* and *B* are entangled, the measurement should change the state of *B* as well. If we now measure $p_B$ we should certainly get $p_B = -p_A$ while the uncertainty in the location of *B* becomes infinite, or very large. This is a substantial change in the wave function $\Psi_B$ which prior to the measurement on *A* could be presented by a wave packet in which both $\Delta x_B$ and $\Delta p_B$ were finite.

The question is, how is it possible that an operation made on *A* instantly influences the state of *B* which, in principle, could be thousands of kilometers away from *A*. The effect of the measurement on *A*, on the state of *B*, is immediate, since the time interval $\Delta t$ between the measurements on *A* and on *B* can be arbitrarily short. It seems to contradict SR which claims that no interaction can travel faster than the speed of light.

## 2. The Spin Version of the EPR paradox

An alternative version of the EPR paradox is based on measuring the spin direction instead of momentum and location (Bohm & Aharonov, 1957). In this version, *A* and *B* are particles with spin 1/2, which were created with opposite spins (e.g. by the decay of a particle with spin 0). They arrive at two detectors $D_A$ and $D_B$ where their spin directions are measured with respect to an arbitrary z-axis. Due to the conservation of angular momentum, if the spin of *A* is found to be positive ($|+½\rangle$), then the spin of the *B* must be negative ($|–½\rangle$), and vice versa. This is true unrelated to the distance that separates the particles at the moment of the measurement.

If we repeat the measurement but measure both spins along the *x*-axis (perpendicularly to the z-axis), the result will be the same. If the spin of *A* is found to be +½, the measured spin of *B* will be –½, and vice versa. The results can be explained in two different ways.

1. The particles are entangled in such a way that when one spin is measured, the other spin becomes its opposite. The quantum state of the system, which could be a combination of $|+½\rangle$ and $|–½\rangle$ for both particles prior to the measurement, collapses into $|+½\rangle$ for particle *A* and $|–½\rangle$ for particle *B* (or vice versa) because of the measurement. This is the QM or the Copenhagen explanation.
2. The two particles were created with definite (opposite) spins around any axis in space we may choose. This is "the hidden variable" explanation.

Both explanations seem to be problematic. In order to illustrate the problem with the first explanation, let's assume that particle *A* arrives at the detector $D_A$ and after a short time interval, $\Delta t$, particle *B* arrives at the detector $D_B$. The information about the measured spin of *A* should travel from $D_A$ to $D_B$ faster than the speed of light, since the distance between them can be many kilometers, while $\Delta t$ can be arbitrary small. This seems to violate SR.

According to the second explanation, no information is transferred from one detector to the other. However, we have to assume that each of the two particles is created with definite eigenvalues of both $S_z$ and $S_x$. But, according to QM, two perpendicular components of the angular momentum cannot be simultaneously in defined states. If a particle has a definite spin direction relative to the *z* axis, its spin direction on the *x* axis should be a superposition of $|+½\rangle$ and $|-½\rangle$ so that a measurement can give each state with equal probabilities. If the information about the results of potential measurements of the spin along any arbitrary axis exists prior to the measurement, that information must somehow be concealed. Hence this model was dubbed "hidden variables."

## 3. Bell's Experimental Setup for Photons

Until the article of John Stewart Bell (1964), no experiment was made in order to find out which of the two explanations is true, since they were thought to be indistinguishable in terms of experimental results. Bell pointed out that the two explanations predict the same results if $D_A$ and $D_B$ are oriented in the same direction in space. However, the results predicted by the two explanations may differ, if the detectors measure the spins in different directions. Bell adopted the spin example advocated by Bohm and Aharonov (1957), but since most experiments were made with photons (where the polarization of the photon substitutes the spin component), it is be more convenient in the ensuing discussion to present Bell's thought experiment with photons. The following scheme, depicted in Fig 1, is a simplified version of the two-channel experimental setup first employed by Aspect et al. (1982).

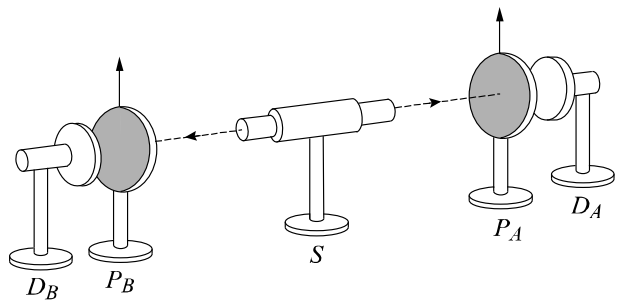

Figure 1

A source (*S*) emits two photons, *A* and *B*, which travel in opposite directions and have the same polarization state. The photons reach two detectors (e.g., sensitive photomultipliers), $D_A$ and $D_B$ which are able to detect single photons. In front of each detector, there is a polarizer ($P_A$ and $P_B$). If the photon passes the polarizer and arrives at the detector, the event is registered as 1. If the photon is stopped by the polarizer and does not reach the detector, the event is recorded as 0.

Let's assume first that the axes of both polarizers are oriented in the *z* direction as shown in Fig. 1. Since the two photons were emitted with the same polarization, we can assume that there will be perfect correspondence between the results recorded by the two detectors: if one detector records the series: 1, 0, 0, 1, 0, 1..., the second detector will record exactly the same series. On the other hand, if polarizer $P_A$ is oriented in the *z* direction while polarizer $P_B$ is oriented in the *x* direction, there will be a perfect mismatch between the results: if $D_A$ records 1 then $D_B$ will record 0, and vice versa.

Suppose that there is an angle $\theta$ between the two polarizers. We define a "matching function" $F(\theta)$ which is the ratio of the number of matches between the two detectors and the total number of readings, in a long series of measurements. We also define a "mismatch function" $E(\theta)$ as the percentage of mismatches between the two detectors. It's easy to see that:

(1) $E(\theta) = 1 - F(\theta)$ (for any $\theta$)
(2) $F(\theta = 0^0) = 1$ ; $E(\theta = 0^0) = 0$
(3) $F(\theta = 90^0) = 0$ ; $E(\theta = 90^0) = 1$

Let's consider the case of $\theta = 0^0$. The correlation between $D_A$ and $D_B$ (Eq. 2) comes as no surprise if the initial polarization of the pair of photons is parallel to the *z* axis ($|z\rangle$) or to the x axis ($|x\rangle$). However, we expect to get the same correlation even when the initial polarization of the photons form an arbitrary angle $\phi$ ($0^0 < \phi < 90^0$) with the z axis. In this case, the initial polarization prior to the measurement can be considered as a superposition of $|z\rangle$ and $|x\rangle$, and the measurement actually causes the collapse of the wave function of each photon to one of the eigenstates, $|z\rangle$ or $|x\rangle$. As mentioned above, the fact that in two remote locations the collapse is to the same eigenstate can be explained in two alternative ways: the QM interpretation (which involves an immediate action at a distance) or the hidden variables model.

## 4. Bell's Inequality and Bell's Theorem

Let's explore the following four-stage gedanken-experiment which will lead us to Bell's Inequality.

Stage 1: $P_A$ and $P_B$ both point in the z direction ($\theta=0$) as shown in Fig. 2-1. Two entangled photons are emitted from the source. Photon A arrives at $P_A$ while photon B arrives at $P_B$. The readings of the two detectors will be the same all the time, therefore $E(\theta)=0$.

Stage 2: $P_B$ is rotated counterclockwise at an angle $\theta$ ($0^0<\theta<45^0$, e.g., $\theta=15^0$) as shown in Fig. 2-2. Once more two photons are emitted from the source. Let's assume that photon A passes through $P_A$ and reaches $D_A$. The probability that photon B will reach $D_B$ is no longer 100%; sometimes it will arrive at $D_B$ and sometimes it will not. In some cases, photon A will be blocked while photon B arrives at $D_B$. Therefore $E(\theta)$, which corresponds to the average mismatch between $D_A$ and $D_B$, is no longer 0.

Stage 3: $P_B$ is restored to its previous position (parallel to the z axis) while $P_A$ is rotated clockwise at the same angle $\theta$, as shown in Fig. 2-3. The situation is symmetrical to Fig. 2-2, and we expect the average mismatch to be $E(\theta)$ as in Stage 2.

Stage 4: We leave $P_A$ as in Stage 3, i.e., skewed at an angle $\theta$ clockwise. We rotate $P_B$ anticlockwise at the same angle $\theta$. The angle between the two polarizers is now $2\theta$, as shown in Fig. 2-4. We repeat the series of measurements that were performed in the previous stages. The mismatch between $D_A$ and $D_B$ is now $E(2\theta)$.

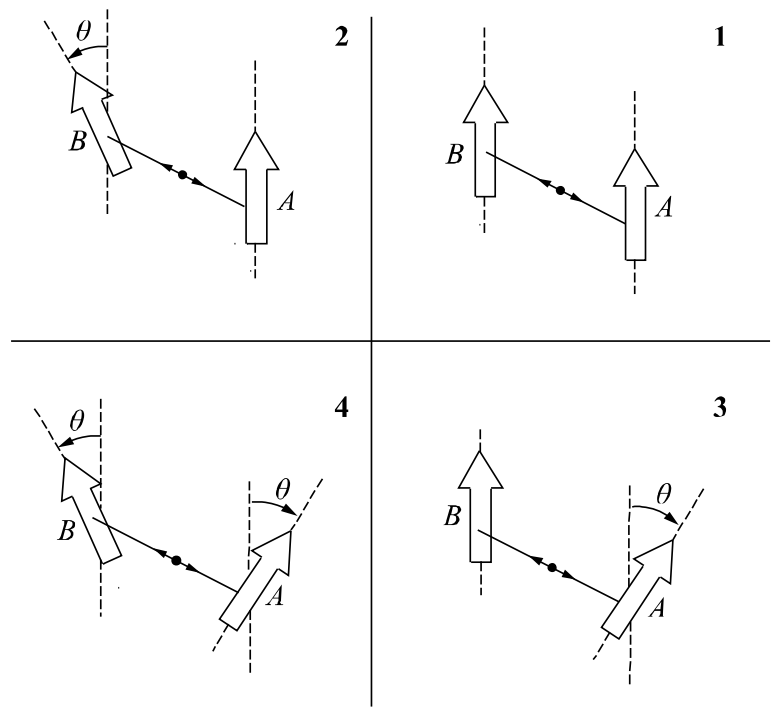

**Figure 2**

We shall now see that the two models which explain the correlation between the readings of $D_A$ and the readings of $D_B$ provide different predictions about the relationship between $E(2\theta)$ and $E(\theta)$. Therefore, by measuring $E(\theta)$ and $E(2\theta)$ for various values of $\theta$, it can be determined which of the two models is correct.

According to the hidden variable model – that rejects action at a distance – there is no connection between the two polarizers. There is no way that polarizer $P_A$ could sense the state of polarizer $P_B$. Therefore, the rotation of polarizer $P_B$ between Stage 3 and Stage 4 cannot affect polarizer $P_A$. Therefore, at Stage 4, polarizer $P_A$ will continue to pass or block, photons with the same rate of mismatch which we would get if $P_B$ stayed upright. In other words, at Stage 4, the mismatch between the series *a* (which is the readings of $D_A$) and a hypothetical series *c* (which represents the results we would get from $D_B$ at Stage 4, if $P_B$ remained upright) is still $E(\theta)$.

By the same token, the mismatch between *c* and *b* (the actual readings of $D_B$ at Stage 4) will also be $E(\theta)$. One might conclude that the mismatch between *a* and *b* will be the sum of the mismatch between *a* and *c* and the mismatch between *b* and *c*, namely, that $E(2\theta)=2E(\theta)$. However, this conclusion is too hasty. In some cases, an element in *a* will be the opposite of the corresponding element in *c*, and the same element in *b* will also be the opposite of the element in *c*. In this case, the elements in *a* and in *b* will be identical. Therefore, the average mismatch between *a* and *b* at Stage 4, namely $E(2\theta)$, can be $2E(\theta)$ but can also be smaller than $2E(\theta)$. We can write:

(4) $\quad E(2\theta) \leq 2E(\theta) \quad$ [Bell's inequality]

We can demonstrate the relation between $E(2\theta)$ and $2E(\theta)$ by the following example. Let's assume that

(5) $\quad c = 0, 0, 1, 0, 1, 0, 1, 1, 0, 1, 1, 0$

In order to create series *a,* we duplicate *c* but randomly change 4 of the 12 digits. To create series *b,* we once more duplicate *c* and randomly change 4 of the 12 digits. The mismatch between *c* and *a* as well as between *c* and *b* will be 1/3.

(6) $\quad E(c,a) = E(c,b) = 1/3$

If the elements of *c* which were changed to create *b* were different from the elements of *c* which were changed to create *a*, the mismatch between *a* and *b* will be 2/3. However, if the same elements were changed in both cases, the mismatch between *a* and *b* will be 0. In the general case we can write:

(7) $\quad E(a,b) \leq 2/3$

In correspondence with Eq. (4). Equation (4) is Bell's inequality for the specific experimental procedure described above. According to Bell's theorem, Eq. (4) would be verified experimentally if the hidden variables model is true. On the other hand, if Bell's inequality is violated, then the QM model is true.

## 5. $E(\theta)$ According to Quantum Mechanics

We can go on and evaluate $E(\theta)$ according to the traditional interpretation of QM. Let's assume that in Stage 2 of the thought experiment described in Section 4, photon *A* passes through $P_A$ and reaches $D_A$. This means that the polarization of photon *A* is parallel to the *z* axis. Since the two photons are entangled, this is also the polarization of photon *B*. The angle between the axes of $P_A$ and $P_B$ is $\theta$. Therefore, the polarization of photon *B* is skewed relatively to $P_B$ at an angle $\theta$.

According to Malus' law, when a polarized beam of light hits a perfect polarizer, the intensity of the light that passes through the polarizer is given by:

(8) $\quad I = I_0 \cos^2 \theta$

where $I_0$ is the initial intensity and $\theta$ is the angle between the light's direction of polarization and the axis of the polarizer.

When we regard the beam as a stream of photons, we can ascribe to each photon a defined direction of polarization which creates an angle $\theta$ with $P_B$. According to Eq. (8), out of *N* photons, approximately $N\cos^2\theta$ will pass $P_B$ and about $N\sin^2\theta$ will be blocked. The probability that a single photon will reach $D_A$ or $D_B$, while its companion will be blocked at the other detector, is therefore $\sin^2\theta$, and this will be the average mismatch rate in a long series of measurements[3]. Thus, according to QM, at Stages 2 and 3:

(9) $\quad E(\theta) = \sin^2 \theta \qquad$ (Stages 2 and 3)

In Stage 4, the angle between the polarizers is $2\theta$ and, according to QM, the mismatch rate will be:

(10) $\quad E(2\theta) = \sin^2 2\theta$

If, for example, $\theta = 30°$, we shall get in Stages 2 and 3 a mismatch rate of:

(11) $\quad E(30°) = \sin^2(30°) = 0.5^2 = 0.25$,

while in Stage 4 the mismatch will be:

(12) $\quad E(60°) = \sin^2(60°) = \left(\dfrac{\sqrt{3}}{2}\right)^2 = 0.75$

---

[3] If photon *A* is blocked, the probability that photon *B* will pass is $\cos^2(90°-\theta)=\sin^2\theta$

In this case $E(2\theta) > 2E(\theta)$ in contrast to Bell's inequality (Eq. 4). Indeed, it's easy to prove that $E(2\theta) > 2E(\theta)$ for any $\theta$ in the range: $0° < \theta < 45°$. To prove this, we notice that:

(13) $\quad \sin^2(2\theta) = 4\sin^2(\theta) \times \cos^2(\theta)$

In the range $0° < \theta < 45°$, the function $\cos^2(\theta)$ is a monotonically descending function which has a minimum at $\theta = 45°$, where $\cos^2(\theta) = 0.5$. At that point:

(14) $\quad \sin^2(2\theta) = 4\sin^2(\theta) \times 0.5 = 2\sin^2(\theta) \quad (\theta = 45°)$

namely, for $\theta = 45°$, $E(2\theta) = 2E(\theta)$. If $0° < \theta < 45°$ then $\cos^2\theta > 0.5$ and $\sin^2(2\theta) > 2\sin^2(\theta)$ which means that $E(2\theta) > 2E(\theta)$, in contrast with Bell's inequality.

Fig. 3 depicts $E(\theta)$ according to the QM description ($E(\theta) = \sin^2\theta$) in the range $0° \leq \theta \leq 90°$. On the straight line (described by $E(\theta) = \theta/90°$) the relation between $E(\theta)$ and $E(2\theta)$ is: $E(2\theta) = 2E(\theta)$. One can see that in the range $0° < \theta < 45°$, the function $E(\theta) = \sin^2\theta$ (which represents QM's results) lies below the straight line, while Bell's inequality (Eq. 4) is valid only for points which are above the straight line. Thus, by performing the experiment described in Sec. 4, one can find out whether the QM model or the hidden variables model is correct.

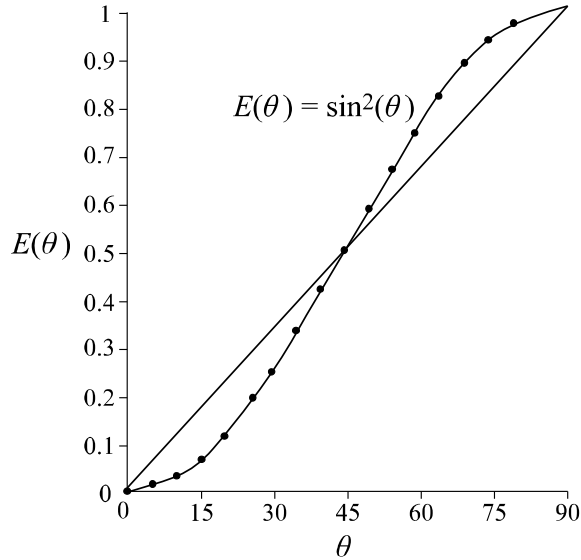

**Figure 3**

In a pioneering work of Freedman and Clauser (1972), an experiment similar to the gedanken-experiment described above was performed. They measured the polarization correlation of two entangled photons emitted in an atomic cascade of calcium. The wavelengths of the photons were 551.3 nm and 422.7 nm. The measurements were made at nine different angles in the range $0° \leq \theta \leq 90°$. The results were in agreement with quantum mechanics, and violated

Bell's inequality to a high statistical accuracy. Actually, the curve describing the results (Fig. 3 in Freedman & Clauser, 1972) is similar to $E(\theta) = \sin^2\theta$ in Fig 3 above, except for the values on the vertical axis which reflect the fact that the efficiency of the detectors was less than 100%. The authors considered the results as strong evidence against local hidden-variable theories.

Over the years, additional experiments demonstrated clearly the violation of Bell's inequality. It was established that in experiments such as those described in the EPR and Bell's papers, a measurement performed on one particle does affect the other particle which can be far away. The effect is immediate, due to the fact that a single unified wave function continues to describe the two particles even when they are far apart. Thus, a measurement performed on one particle causes the collapse of the wave function in the other one as well. In order to resolve the contradiction between this action at a distance and SR, we have to see why the contradiction arises in the first place.

## 6. Special Relativity and Action at a Distance

Let $S$ and $S'$ denote two inertial reference frames. $S'$ is moving with respect to $S$ at a constant velocity $\upsilon$ along the $x$-axis. At time $t=0$ the spatial axes and clocks of $S$ and $S'$ coincide. Suppose that an event that occurred in $S$, at point $x_1$ in time $t_1$ creates a signal that travels at a velocity $u$. The signal arrives at a point $x_2$ in time $t_2$, and creates a second event there (e.g., turning on a lamp). The coordinates and times of the two events in $S'$ are $(x'_1, t'_1)$ and $(x'_2, t'_2)$. According to the Lorentz transformation:

$$(15) \quad t'_1 = \gamma\left(t_1 - \frac{\upsilon x_1}{c^2}\right) \quad ; \quad t'_2 = \gamma\left(t_2 - \frac{\upsilon x_2}{c^2}\right)$$

Where $\gamma = 1/(1-\upsilon^2/c^2)^{1/2}$. From (15) we get:

$$(16) \quad \Delta t' = \gamma(\Delta t - \upsilon \Delta x/c^2),$$

Where $\Delta t' = t'_2 - t'_1$ ; $\Delta t = t_2 - t_1$ ; $\Delta x = x_2 - x_1$. Since $\Delta x = u\Delta t$, we can write:

$$(17) \quad \Delta t' = \gamma \Delta t (1 - u\upsilon/c^2)$$

If the signal velocity $u$ is greater than the speed of light ($u>c$), we can define a reference frame $S'$ that moves at a velocity $\upsilon$ which is smaller than $c$ but close enough to $c$ so that $u\upsilon/c^2 > 1$ (in order for that to happen, $\upsilon$ should meet the condition $c > \upsilon > c^2/u$). The expression $(1 - u\upsilon/c^2)$ in Eq. (17) will be negative, and therefore if $\Delta t$ is positive, $\Delta t'$ should be negative and vice versa. Thus, in $S'$, event 1 will occur after event 2 and an observer in $S'$ would see the effect precede its cause. For example, let $u = 1.1c$ and $\Delta x = c \times 1$s while $\upsilon = 0.98c$ ($\gamma = 5.025$). With these values:

$\Delta t = \Delta x/u = 0.9091$ s

$\Delta t' = 5.025 \times 0.9091(1 - 0.98 \times 1.1) = -0.356$ s

The minus sign indicates that the order of events in $S$ and $S'$ is reversed . According to an observer in $S'$, event 1 (which is the cause of event 2) occurred 0.356 seconds **after** event 2. The assumption that a signal can travel faster than the speed of light leads to a violation of the relativistic principle of causality, which asserts that an effect never precedes its cause, in any reference frame. This is why SR forbids instantaneous action at a distance as well as travelling of matter, energy or information at speeds greater than the speed of light.

In order to demonstrate the inconsistency of Bell-like experiments with SR, let's return to Stage 1 in the experiment described in Section 4. Both polarizers $P_A$ and $P_B$ are oriented in the $z$ direction ($\theta=0$), as shown in Fig. 2-1. Thus the readings of the two detectors are the same all the time and $E(\theta) = 0$. Let's assume that $D_A$, $D_B$ and the source are on the $x$ axis of a rest frame $S$. The source is at $x_0=0$ and the coordinates of $D_A$ and $D_B$ are:

(18)    $x_1 = x(D_A) = 15$m ;   $x_2 = x(D_B) = -15.3$ m

In order to facilitate the calculations, we redefine our unit of length so that $c=3\times10^8$m exactly. We denote by $t_1$ and $t_2$ the times of two events. **Event 1**: "Photon $A$ reached $P_A$ and then $D_A$ registered 0 or 1". **Event 2**: "Photon $B$ reached $P_B$ and then $D_B$ registered 0 or 1". It's easy to see that:

(19)    $t_1 = 50$ ns ; $t_2 = 51$ ns

Before checking the times in another frame, $S'$, let's discuss the following question: Can an observer in $S$ consider event 2 as the result of event 1? I claim that she can. The two events can be considered a combination of cause and effect for the following reasons.

1.  In frame $S$, event 1 precedes event 2 by 1 ns.

2. Prior to the occurrence of event 1, the reading of $D_B$ could be either 0 or 1 in equal probabilities. After event 1 takes place, the reading of $D_B$ is definitely determined.

3. If $D_A$ was removed after the photons left the source, and before they reached the polarizers, the reading of $D_B$ could be either 0 or 1 in equal probabilities. The fact that $D_A$ operated and registered the arriving photon influenced the reading of $D_B$.

Therefore, event 2 can be considered the result of event 1. An argument against this claim is that the two events are separated by a spacelike interval. According to SR, only if two events are separated by a timelike or lightlike interval, can one event influence the other. However, it can be claimed that the concept of causality is metaphysically prior to the relativistic restrictions. Actually, it was argued that many standard philosophical theories would treat the relationship between such two events as causal despite the contradiction with SR (Butterfield 1992).

Let's assume a second frame $S'$ which is moving with respect to $S$ in the negative direction of the $x$ axis at a velocity $\upsilon = -0.6c$. Let's assume that when the two photons are ejected from the source, the clocks at $S$ and $S'$ show $t = t' = 0$, and the origins coincide. By using the Lorentz transformation (Eq. 15) we find:

(20)   $t'_1 = 100$ ns ; $t'_2 = 25.5$ ns

Thus, in $S'$ event 2 occurs **before** event 1, although we defined event 1 as the cause of event 2. Actually, since the velocity of the signal which carries the information between $D_A$ and $D_B$ is infinite, the paradox appears for $\upsilon$ as small as $|\upsilon| \approx 0.01c$ for the numerical values of the example above. In general, it occurs whenever:

(21)   $-\upsilon \geq \dfrac{t_2 - t_1}{x_1 - x_2} c^2$

## 7. Solutions to the Paradox

Several solutions can be offered for the contradiction which was demonstrated in the previous section between SR and Bell-like experiments. Ballentine and Jarrett (1987) suggested a distinction between a "strong" locality principle and a "weak" one that is needed to satisfy the demands of relativity. They claimed that QM satisfies the latter and therefore there is no contradiction between QM and SR. Instead, one can argue that we do not have two separate events here but only one spatially extended but indivisible event which is "the collapse of the wave-function which represents the polarization of the two photons" (Butterfield, 1992, p. 41). Another alternative is to formulate a theory of causation which requires some conditions which two events need to fulfil in order to represent a cause and effect relationship, and then show that these conditions are not realized here (ibid, p. 42).

The solution which I suggest to the paradox is based on the following principles:

1. Event 1 and event 2, in the example discussed in Section 6, are two distinct and separate events which occur at different points in space-time.
2. There is a cause and effect connection between the two events.
3. According to an observer in $S$, event 2 is a result of event 1.
4. According to an observer in $S'$, on the other hand, event 1 is a result of event 2.
5. The disagreement between the two observers does not violate the causality principle of SR, since in this particular case the cause and effect relationship between the two events is symmetrical: each of them can be regarded as a result of the other, depending on the frame of reference in which they are observed.

Usually, when there is a causal relationship between two remote events, they are physically different. That's why the effect cannot precede its cause in any reference frame. For example, if event 1 is the ejection of a signal from point ($x_1$, $y_1$, $z_1$, $t_1$) and event 2 is the arrival of the signal to point ($x_2$, $y_2$, $z_2$, $t_2$) where it turns on a lamp, we demand that, in any reference frame, $t_2 > t_1$. This demand is fulfilled only if the velocity of the signal does not exceed the speed of light, as shown in Section 6.

However, in Bell-like experiments, like the one described in Section 4, there is no physical difference between the two events: they are totally symmetrical. Each of them can serve as a cause or as an effect, depending on the frame of reference in which they are observed. If event 1 is observed before event 2, event 1 is the cause and event 2 is the result. If the order of times is reversed, then event 2 is the cause and event 1 is the result.

It is customary to think that the causal relation between two events can be one of three types:

1. Event 2 is a result of event 1.

2. Event 1 is a result of event 2.

3. There is no causal relation between the two events.

The analysis of the ostensible contradiction between Bell's theorem and SR indicates that there is a fourth possibility. In entangled systems, one can find pairs of "entangled events" which have symmetrical cause and effect relations. Each of them can appear to be the cause of the other, depending on the frame of reference in which they are observed. This fourth possibility solves the paradox which the action at a distance creates.